\documentstyle[12pt]{article}
\begin{document}
\begin{center}
{\large \bf Effective Neutrino Mass Operators: \\
\smallskip
A Guide to Model Building}\footnote{Contribution to the
NuFact'01 Workshop Proceedings.} \\
\vspace{1.0 cm}
C. N. Leung\footnote{Supported in part by the U.S. Department
of Energy under grant DE-FG02-84ER40163.} \\
\bigskip
Department of Physics and Astronomy, University of Delaware\\
\smallskip
Newark, DE 19716, U.S.A. \\
\vspace{1.0 cm}

{\bf Abstract}
\end{center}
Effective operators relevant for generating small Majorana
masses for the neutrinos in the Standard Model will be
considered.  These operators serve as a useful guide for
building models of neutrino mass.  Some of these operators
are represented by familiar models in the literature, and
others lead to interesting new models.  The number of the
relevant operators will be drastically reduced if neutrinoless
double beta decays are observed in current experiments.

\newpage

I would like to report some recent work with K.S.
Babu\cite{BuL}.  In this work, we compiled all the effective
higher dimensional ($5 \le d \le 11$) operators (and considered
renormalizable models that can produce them) which can generate
small Majorana neutrino masses consistent with that indicated
by neutrino oscillation experiments.  As you know, data from
these experiments suggest a neutrino spectrum with masses of
order 1 eV or less.  It is one of the challenges of the fermion
mass problem to understand why the neutrinos are so much
lighter than the charged leptons and quarks.

A natural way to comprehend the smallness of the neutrino mass is
to assume that they are generated via some underlying new physics
at an energy scale $\Lambda$ which is higher than the electroweak
scale.  Typically, $\Lambda$ corresponds to the scale at which the
lepton number symmetry is broken.  At or below the electroweak
scale, the neutrino mass will then be described by higher
dimensional ($d > 4$) effective operators which are suppressed by
appropriate powers of $\Lambda$.  One can then understand the
smallness of the neutrino mass on purely dimensional grounds,
without having to know exactly what the underlying new physics is.

An example of this effective--operator description is the
well--known seesaw mechanism~\cite{seesaw} in which heavy
$SU(2)_L$ singlet right--handed neutrinos, $N$, are introduced
which can form a Dirac mass term with the left--handed neutrinos.
Upon integrating out the heavy $N$ fields, one obtains an
effective theory without the right--handed neutrinos, but with
the dimension 5 operators~\cite{weinberg} (see Ref.\cite{BuL}
for an explanation of the notation): ${\cal O}_1 = (L_a^{iT}
C L_b^j) H^k H^l \epsilon_{ik} \epsilon_{jl}$, which can
generate small Majorana masses for the left--handed
neutrinos.  When the neutral component of $H$ develops its
vacuum expectation value, $v$, ${\cal O}_1$ will produce a
Majorana mass matrix for the neutrinos, with eigenvalues of
order $v^2/\Lambda$.  In order to generate a neutrino mass of
order 1 eV or less, $\Lambda$ must be greater than $10^{13}$
GeV or so.

On the other hand, the nonobservation of lepton number violating
processes such as $\mu \rightarrow e e e$ and $\mu \rightarrow e
\gamma$ only constrains $\Lambda$ to be larger than a few TeV.
If the actual lepton number breaking scale is closer to this
experimental lower limit\footnote{Such may be the case, e.g.,
if lepton number is broken by quantum effects of gravity.  Then
$\Lambda$ will be of order the Planck scale which can be as low
as a few TeV if large extra dimensions exist.}, ${\cal O}_1$
will generate too large a neutrino mass and other effective
neutrino mass operators will have to be considered.  In view
of the current interests in neutrino mass models, it will be
useful to identify all such operators.

Since we are interested in effective operators that can lead to
a Majorana mass term for the left-handed neutrino fields in the
Standard Model, the operators must violate lepton number, L, by
two units.  They must be $SU(3)_C \times SU(2)_L \times U(1)_Y$
invariant and are also required to conserve baryon
number\footnote{Otherwise, the nonobservation of proton decays
will limit $\Lambda$ to be larger than $10^{14}$ GeV and the
resulting neutrino masses will be too small to be of interest.},
B.  Thus, these operators violate (B - L) and may be relevant
for models of leptogenesis\cite{leptogen}.

The full list of effective neutrino mass operators can
be found in Ref.\cite{BuL}.  It is also shown there how to
construct neutrino mass models systematically from these
operators.  We classify the effective operators according
to the number of fermion fields they contain.  Operators
containing two fermion fields are just the $d = 5$ seesaw
operators given by ${\cal O}_1$, which generate neutrino
masses at tree level.

Operators containing four fermion fields will generate neutrino
masses radiatively at the one--loop level.  Examples of this
class of operators include ${\cal O}_2 = L^i L^j L^k e^c H^l
\epsilon_{ij} \epsilon_{kl}$, which have a realization in the
Zee model~\cite{zee}, and ${\cal O}_3 = L^i L^j Q^k d^c H^l
\epsilon_{ik} \epsilon_{jl}$, which are realized in the
supersymmetric standard model with $R$-parity violation\cite{hall}.
As an example of how to use the effective operators to obtain
new neutrino mass models, consider the operator ${\cal O}_4 =
L^i L^j \bar{Q}_i \bar{u^c} H^k \epsilon_{jk}$, which can
generate one--loop neutrino masses via an $SO(10)$ grand unified
model (see Ref.\cite{BuL} for details).

Operators containing six fermion fields will generate neutrino
masses as two--loop radiative corrections.  Because of the
additional suppression factors, the scale $\Lambda$ in this
class of models will be much lower than the seesaw scale, and
may even be close to the electroweak scale.

We have not considered operators with eight or more fermion
fields because operators with $d \ge 12$ will be highly
suppressed and will generate neutrino masses that are too
small to satisfy the atmospheric neutrino data which require
at least one neutrino to have a mass of about 0.03 eV.  Even
with this truncation, our list contains a large number of
operators.  It will be helpful if one can find a way to reduce
the number of relevant operators.  Neutrinoless double beta
($2 \beta 0 \nu$) decays may provide such a way.  This is because
there is a subset of operators (e.g., $L^i L^j Q^k d^c Q^l d^c
\epsilon_{ik} \epsilon_{jl}$, $L^i d^c \bar{Q}_j \bar{u^c}
\bar{e^c} \bar{u^c} \bar{H}_i H^j$, etc.) which have the special
property that they can produce $2 \beta 0 \nu$ decays directly,
but can only generate neutrino masses at the two--loop level.
These operators can therefore generate $2 \beta 0 \nu$ decay
amplitudes which are large enough to be observable in current
experiments even though the neutrino masses they induce are as
small as that indicated by the solar and atmospheric neutrino
data.  Thus, if $2 \beta 0 \nu$ decays are observed in the
current round of experiments, this subset of operators will be
singled out as the most likely effective neutrino mass operators.

\end{document}